\documentclass[12pt]{article}
\usepackage{pstricks}
\usepackage{longtable}
\usepackage[dvips]{graphicx}       
\usepackage{color}
\usepackage{cite}
\usepackage{amssymb}
\input{epsf} 
\def\beq{\begin{equation}} 
\def\eeq{\end{equation}}

\def\b0{\beta_0}

\def\beeq{\begin{eqnarray}}
\def\eeeq{\end{eqnarray}}

\def\mur{\mu_R} 
\def\muf{\mu_F}
\def\mur2{\mu_R^2} 
\def\muf2{\mu_F^2}

\newcommand {\apgt} {\ {\raise-.5ex\hbox{$\buildrel>\over\sim$}}\ }
\newcommand {\aplt} {\ {\raise-.5ex\hbox{$\buildrel<\over\sim$}}\ }

\begin{document} 

\begin{titlepage}
\renewcommand{\thefootnote}{\fnsymbol{footnote}}
\begin{flushright}
\end{flushright}
\par \vspace{-5mm}

\begin{center}
{\Large \bf

Quark and gluon spin-2 form factors to two-loops in QCD

}
\end{center}
\par \vspace{-1mm}
\begin{center}
{\bf Daniel de Florian}\footnote{deflo@df.uba.ar}$^{(a)}$,
{\bf Maguni Mahakhud}\footnote{maguni@hri.res.in}$^{(b)}$,
{\bf Prakash Mathews}\footnote{prakash.mathews@saha.ac.in}$^{(c)}$,\\
[5pt]
{\bf Javier Mazzitelli}\footnote{jmazzi@df.uba.ar}$^{(a)}$ and
{\bf V. Ravindran}\footnote{ravindra@imsc.res.in}$^{(d)}$\\

\vspace{10mm}

$^{(a)}$ Departamento de F\'\i sica, FCEyN, Universidad de Buenos Aires, \\
(1428) Pabell\'on 1, Ciudad Universitaria, Capital Federal, Argentina

$^{(b)}$ 
Regional Centre for Accelerator-based Particle Physics\\
Harish-Chandra Research Institute, Chhatnag Road, Jhunsi,\\
Allahabad 211 019, India

$^{(c)}$ Saha Institute of Nuclear Physics, 1/AF Bidhan Nagar, Kolkata 700 064, India

$^{(d)}$ The Institute of Mathematical Sciences\\
C.I.T Campus, 4th Cross St, Tharamani Chennai,\\
Tamil Nadu 600 113, India

\vspace{5mm}

\end{center}

\par \vspace{2mm}
\begin{center} {\large \bf Abstract} \end{center}
\begin{quote}
\pretolerance 10000

We present complete two-loop radiative corrections to the graviton-quark-antiquark
form factor $G^* \rightarrow q~\overline q$ and graviton-gluon-gluon form factor
$G^* \rightarrow g~g$ in $SU(N)$ gauge theory with $n_f$ light flavours using 
$d$-dimensional regularisation to all orders in $\varepsilon=d-4$.  This is an
important ingredient to next-to-next-to-leading order QCD corrections to hadronic
scattering processes in models with large extra-dimensions where Kaluza-Klein
graviton modes couple to Standard Model fields.  We show that these form factors
obey Sudakov integro-differential equation and the resulting cusp, collinear and
soft anomalous dimensions coincide with those of electroweak vector boson and gluon
form factors.  We also find the universal behaviour of the infrared singularities
in accordance with the proposal by Catani.

\end{quote}

\vspace*{\fill}
\begin{flushleft}
\end{flushleft}
\end{titlepage}

\setcounter{footnote}{1}
\renewcommand{\thefootnote}{\fnsymbol{footnote}}

\section{Introduction}

Extra dimension models are interesting scenarios to tackle the hierarchy
problem.  Depending on the geometry of the extra dimension(s), the two
popular options are: (a) flat extra dimension model (ADD) \cite{ADD} and
(b) the warped extra dimension model (RS) with a large curvature \cite{RS1}.
In both these models, only the graviton is allowed to permeate the bulk
which leads to different spectrum of spin-2 Kaluza-Klein (KK) modes in
4-dimensions.  The spin-2, KK modes couple to the Standard Model (SM)
particles through the energy momentum tensor of the SM.

These beyond SM model scenarios could alter the SM predictions by additional
virtual KK mode exchanges and real KK mode productions.  Dedicated groups in
both ATLAS \cite{atlas} and CMS \cite{cms} collaborations are engaged in the
analysis for extra dimension searches in various processes like di-lepton,
di-photon, mono-jet, mono-photon productions etc.  To put stringent bounds
on the parameters of these BSM models, control on the theoretical uncertainties
is essential.  Renormalisation and factorisation scale dependences of a cross
section to a particular order in perturbation theory give an estimate of the
uncalculated higher order corrections.  Presently next-to-leading order (NLO)
QCD calculations have been done for di-lepton \cite{di-ll}, di-photon
\cite{di-ph} and di-electroweak gauge boson \cite{di-ZZ} productions via virtual
KK modes in addition to the SM contributions.  These virtual contributions
have been incorporated in the {\scriptsize A}MC@NLO frame work and results
to NLO+PS accuracy are now available for most of the di-final state processes
\cite{di-final+ps}.  In all these processes the factorisation scale dependence
reduces substantially and in addition the NLO correction is in fact significant.
For the above processes, the leading order (LO) is of the order ${\cal O} 
(\alpha_s^0)$, the renormalisation scale dependence starts only at NLO in QCD.
To control the renormalisation scale dependence one would have to go to 
next-to-next-to-leading order (NNLO) order.

A full NNLO QCD contribution requires the knowledge of graviton-quark-antiquark
$G^* \rightarrow q \overline q$ and graviton-gluon-gluon $G^* \rightarrow g g$
form factors up to two-loop level in QCD in addition to double real emission and
one-loop single real emission scatter processes at the parton level. Here $G^*$
denotes the virtual graviton. 
In this article we take the first step towards the full NNLO computation by
evaluating these form factors to two-loop level in QCD by sandwiching the
energy momentum tensor of the QCD part of the SM between on-shell gluon and
quark states.  We will also discuss the infrared (IR) structure of these form
factors using Sudakov's integro-differential equation and Catani's proposal
on two-loop QCD amplitudes.

In the next section, we will derive the two-loop form factors and in the 
section 3, we describe the infrared structure of these form factors.  Finally
we conclude in section 4 and the appendix contains the form factors expanded
in powers of $\varepsilon$ to desired accuracy.

\section{Two loop form factors}
We work with the following action that describes the interaction of SM fields
with the KK modes of the gravity.  To lowest order in $\kappa$, the KK modes
couple to SM fields through energy momentum tensor of SM.  Here, we restrict
ourselves to QCD part of the energy momentum tensor:
\begin{eqnarray}
{\cal S} = {\cal S}_{SM} - {\kappa \over 2} \int d^4 x ~T^{QCD}_{\mu\nu}
(x)~ h^{\mu\nu} (x) \, ,
\end{eqnarray}
where $T^{QCD}_{\mu\nu}$ is the energy momentum tensor of QCD \cite{di-ll}: 
\begin{eqnarray}
T^{QCD}_{\mu\nu} &=& -g_{\mu\nu} {\cal L}_{QCD} - F_{\mu\rho}^a F^{a\rho}_\nu
-{1 \over \xi} g_{\mu\nu} \partial^\rho(A_\rho^a\partial^\sigma A_\sigma^a)
\nonumber\\[2ex]
&&+{1 \over \xi} (A_\nu^a \partial_\mu(\partial^\sigma A_\sigma^a) + A_\mu^a\partial_\nu
(\partial^\sigma A_\sigma^a))
+{i \over 4} \Big[ \overline \psi \gamma_\mu (\overrightarrow{\partial}_\nu -i g_s T^a A^a_\nu)\psi
\nonumber\\[2ex]
&&-\overline \psi (\overleftarrow{\partial}_\nu + i g_s T^a A^a_\nu) \gamma_\mu \psi
+\overline \psi \gamma_\nu (\overrightarrow{\partial}_\mu -i g_s T^a A^a_\mu)\psi
\nonumber\\[2ex]
&&-\overline \psi (\overleftarrow{\partial}_\mu + i g_s T^a A^a_\mu) \gamma_\nu \psi\Big]
+\partial_\mu \overline \omega^a (\partial_\nu \omega^a - g_s f^{abc} A_\nu^c \omega^b)
\nonumber\\[2ex]
&&+\partial_\nu \overline \omega^a (\partial_\mu \omega^a- g_s f^{abc} A_\mu^c \omega^b).
\end{eqnarray}
In the above equation, $g_s$ is the strong coupling constant and $\xi$ is gauge parameter
in Lorenz gauge fixing condition.  The peculiar feature in the above action
is the appearance of the direct coupling of ghost fields ($\omega,\overline \omega$) 
with KK modes \cite{di-ll}.  We have kept track of these unphysical contributions 
along with those coming from gauge fixing term
in order to establish the cancellation of their contributions among themselves.    
We have retained only light flavours in the quark sector.

We compute the relevant form factors by evaluating the truncated matrix 
elements $\hat \mathbf {\cal M}_I$ of $T^{QCD}_{\mu\nu}$
between on-shell gluon ($I=g$) and quark/anti-quark ($I=q,\overline q$) states.
The symbol  $\hat {}$  here and in the following denotes that
the quantities are unrenormalised/bare.
The $\hat \mathbf {\cal M}_I$s in the color space can be expanded
as
\begin{eqnarray}
\mathbf{\hat {\cal M}}_I =\mathbf{\hat {\cal M}}_I^{(0)} + \hat a_s 
\left({Q^2 \over \mu^2}\right)^{{\varepsilon\over 2}} S_\varepsilon \mathbf{ \hat {\cal M}}_I^{(1)} 
+ \hat a_s^2 \left({Q^2 \over \mu^2}\right)^{{\varepsilon}} S_\varepsilon^2 
\mathbf{ \hat {\cal M}}_I^{(2)} + {\cal O}\left(\hat a_s^3\right)\, ,
I=g,q,\overline q \,,
\end{eqnarray}
where the unrenormalised coupling constant $\hat a_s= \hat g_s^2/16 \pi^2$ and
the scale $\mu$ is introduced to keep $g_s$ dimensionless in dimensional
regularisation and the space-time dimension is taken to be $d=4+\varepsilon$.
The scale $Q^2=-q^2 -i \epsilon$, where $q$ is the momentum transfer.  The
 unrenormalised coupling constant $\hat a_s$ is related to the renormalised
one $a_s(\mu_R^2)$ by
\begin{eqnarray}
S_\varepsilon \hat a_s= Z(\mu_R^2) a_s(\mu_R^2) \left({\mu^2 \over \mu_R^2}\right)^{\varepsilon \over 2}
\,,\, \quad \quad S_\varepsilon = \exp\left\{{\varepsilon \over 2} \Big[\gamma_E-\ln 4 \pi\Big]\right\} \,,
\end{eqnarray}
where the renormalisation constant $Z(\mu_R^2)$ is given by
\begin{eqnarray}
Z(\mu_R^2) = 1 + a_s(\mu_R^2) {2 \beta_0 \over \varepsilon} + {\cal O}\left(a_s^2(\mu_R^2)\right)
\,,\,\quad \quad \beta_0={11 \over 3} C_A-{4 \over 3} T_F n_f
\end{eqnarray}
with $\mu_R$-renormalisation scale, $C_A=N,T_F=1/2$ and $n_f$ the number of active flavours.

Using the $\hat \mathbf {\cal M}_I$s, the form factors are defined as
\begin{eqnarray}
\hat F^{T,(n)}_I={\mathbf{\hat {\cal M}}_I^{(0)*}\cdot \mathbf{ \hat {\cal M}}_I^{(n)}  \over
\mathbf{\hat {\cal M}}_I^{(0)*}\cdot\mathbf{\hat {\cal M}}_I^{(0)}} \, ,
\end{eqnarray}
and the symbol $\cdot$ takes care of the color and spin/polarisation sums. 

The Feynman amplitudes that contribute to gluon and quark matrix elements 
of the energy momentum tensor $T^{QCD}_{\mu\nu}$ at born, one-loop and
two-loop levels in QCD are obtained using a computer program QGRAF \cite{qgraf}.  
We find 12 one-loop and 153 two-loop diagrams that contribute to the matrix
element of energy momentum tensor if it is computed between gluon states while
4 one-loop and 54 two-loop diagrams that contribute for quark-antiquark states.  
We have used a set of in-house FORM \cite{Kuipers:2012rf} routines to convert
the QGRAF outputs into a suitable form for further symbolic manipulations.
These FORM routines not only replace the symbolic Feynman vertices, propagators
by the corresponding Feynman rules but also perform Lorentz contractions, Dirac
gamma matrix algebra etc.  We have done all our computations in $d=4+\varepsilon$
dimensions in order to regulate both ultraviolet (UV) and infrared (IR) divergences.  
The resulting expressions at this stage contain one and two-loop tensor and scalar
integrals.  Since the coupling of KK modes with the energy momentum tensor involves
quadratic derivatives, we find that the rank of the tensor integrals present in our
computation is larger than the rank of integrals appearing in quark and gluon form
factors contributing to electroweak vector boson \cite{gon,krla,mane} and Higgs
production cross sections (in the infinite top quark mass limit) 
\cite{harlander,Anastasiou:2002yz,Ravindran:2003um,Ravindran:2004mb} respectively.  

In the past, they were computed using very different methods, that is, different
techniques were employed to perform loop integrals.  In \cite{gon}, the method
of Feynman parameterisation was used in a judicious way so that after each
parametric integration one is left with an integral over the next parameter.
In \cite{krla}, an elegant method, advocated in \cite{chet}, namely ``integration
by parts" (IBP) was used.  
In \cite{mane} the integrals were computed using dispersion techniques
developed in \cite{ne} which uses the Cutkosky
rules \cite{cut}.  In this method, one cuts the Feynman amplitude 
in all possible ways to obtain the imaginary part and the real part
was obtained from the imaginary part via a dispersion relation. 
In \cite{harlander}, an algorithm \cite{basm} which 
relates $l$-loop integrals with $n+1$ external legs to $l+1$-loop integrals with $n$
external legs was used to compute two-loop QCD corrections to gluon form factor 
relevant for Higgs production cross section.
It maps the massless two-loop vertex
functions onto massless three-loop two-point functions which are relatively easy
to compute.
In \cite{Anastasiou:2002yz}, IBP identities were used extensively to compute
the gluon form factor.  
The gluon form factor at two-loop level in $SU(N)$ gauge theory with $n_f$ light flavours 
was computed in 
\cite{Ravindran:2004mb} following \cite{ne}. 
All these results were known only to a desired accuracy in $\varepsilon$, say 
${\cal O}(\varepsilon)$.
In \cite{Gehrmann:2005pd}, using IBP \cite{chet} and Lorentz invariance (LI) \cite{gr} identities, 
the authors
have shown that the two-loop corrections to electroweak quark and gluon form factors 
can be expressed in terms of only few master integrals and have also obtained 
for the first time the closed form solution to one of the master integrals whose
result was known only up to few orders in $\varepsilon$. 
We will closely follow this approach by \cite{Gehrmann:2005pd} to achieve our task.
Reduction of a large number of one and two-loop tensor integrals that appear in
our computation to a few master integrals was 
achieved by FIRE \cite{smirnov}, a Mathematica package, 
which extensively uses the IBP \cite{chet} and 
LI \cite{gr} identities implemented using
Laporta algorithm \cite{laporta}.  Note that there are also similar packages
namely AIR \cite{babis}, Reduce \cite{Studerus:2009ye,vonManteuffel:2012np} and 
most recently LiteRed \cite{Lee:2012cn} that can do this reduction.  
We used LiteRed to cross-check our results obtained using FIRE.  
At one-loop level, we find that the form factors depend only on one master integral 
and at two-loop level, there are one one-loop and three two-loop master integrals.  
These one-loop and two-loop master integrals are now known to all orders in 
$\varepsilon$ and are given in \cite{Gehrmann:2005pd}.   
Below we present our final results for $\hat F_I^{T,(n)}$ for $I=g,q;n=1,2$ 
in terms of these master integrals.

For the gluon form factor, we obtain $\hat F^{T,(0)}_g=1$ and 
\begin{eqnarray}
\hat F^{T,(1)}_g&=& 2 \Bigg[-i A_{2,LO} \Bigg(4 C_A \Big(-68 + 20   ~d + 16   ~d^2 - 8   ~d^3 +   ~d^4\Big) 
\nonumber\\[2ex]
&&   +n_f \Big(-16 + 32   ~d - 15   ~d^2 + 2   ~d^3\Big) \Bigg)\Bigg]\Big/
\Big[(-4 +   ~d) (-2 +   ~d) (2 d^2 - 3 d - 8) \Big]\,, 
\\[2ex]
\hat F^{T,(2)}_g&=&- \Bigg[16 A_{2,LO}^2 \Bigg(-3360 + 5524   ~d - 3607   ~d^2 + 1169   ~d^3 - 188   ~d^4 + 12   ~d^5\Bigg) 
\nonumber\\[2ex]
&&    \times \Bigg\{4 C_A^2 \Big(384 + 3584   ~d - 7712   ~d^2 + 4260   ~d^3 + 128   ~d^4 - 939   ~d^5 + 
       371   ~d^6 
- 62   ~d^7 
\nonumber\\[2ex]
&&
+ 4   ~d^8\Big) + C_F   ~d \Big(1024 - 3616   ~d + 4720   ~d^2 - 
       2926   ~d^3 + 941   ~d^4 
- 153   ~d^5 + 10   ~d^6\Big) n_f 
\nonumber\\[2ex]
&&
  + 8 C_A \Big(192 - 288   ~d - 28   ~d^2 + 322   ~d^3 - 255   ~d^4 
+ 89   ~d^5 - 15   ~d^6 +   ~d^7\Big) n_f\Bigg\} 
\nonumber\\[2ex]
&&
+ A_3 (-8 + 3   ~d) \Bigg\{2 C_A^2 \Big(1720320 
+ 60414976   ~d 
- 195105152   ~d^2 + 
       236351744   ~d^3 
\nonumber\\[2ex]
&& 
- 120445352   ~d^4 
 - 11375804   ~d^5 + 54553314   ~d^6 - 
       36985777   ~d^7 + 13961672   ~d^8 
\nonumber\\[2ex]
&&- 3324848   ~d^9 + 499154   ~d^{10} 
- 43447   ~d^{11} + 1680   ~d^{12}\Big) 
+ 2 C_F   ~d \Big(10379264 
\nonumber\\[2ex]
&&
- 36831232   ~d 
+ 50367872   ~d^2 
- 28580992   ~d^3 
- 3473320   ~d^4 + 16083820   ~d^5 
\nonumber\\[2ex]
&&
- 11518542   ~d^6 + 
       4520247   ~d^7 
- 1098971   ~d^8 + 165551   ~d^9 - 14233   ~d^{10} 
+ 536   ~d^{11}\Big) n_f 
\nonumber\\[2ex]
&&
+ C_A \Big(3440640 
+ 10901504   ~d - 45510400   ~d^2 + 62792448   ~d^3 - 46643440   ~d^4 + 
\nonumber\\[2ex]
&&       20064592   ~d^5 - 4109776   ~d^6 - 494472   ~d^7 + 619031   ~d^8 
- 203281   ~d^9 
+ 36557   ~d^{10} 
\nonumber\\[2ex]
&&
- 3641   ~d^{11} + 158   ~d^{12}\Big) n_f\Bigg\} + 
   2 (12 - 7   ~d +   ~d^2) 
\Bigg(2 A_6 (-4 +   ~d)^2   ~d \Big(-16 + 30   ~d 
\nonumber\\[2ex]
&& 
- 17   ~d^2 + 3   ~d^3\Big) 
      \Bigg\{C_A^2 (3392 
- 3664   ~d + 284   ~d^2 + 794   ~d^3 - 298   ~d^4 + 32   ~d^5) 
\nonumber\\[2ex]
&&  + 2 C_F (-4 +   ~d)^2 
(176 - 26   ~d - 35   ~d^2 + 8   ~d^3) n_f + 
       C_A (-2880 + 1888   ~d + 172   ~d^2 
\nonumber\\[2ex]
&&- 368   ~d^3 + 87   ~d^4 - 6   ~d^5) n_f\Bigg\} + 
     A_4 \Bigg\{2 C_A^2 (-1720320 + 25437184   ~d 
\nonumber\\[2ex]
&&- 55822976   ~d^2 + 41289728   ~d^3 + 
         1696440   ~d^4 - 22168812   ~d^5 
+ 16330266   ~d^6 
\nonumber\\[2ex]
&&
- 6288301   ~d^7 + 
         1498316   ~d^8 - 234230   ~d^9 + 24945   ~d^{10} 
- 1812   ~d^{11} + 72   ~d^{12}) 
\nonumber\\[2ex]
&&
+ 2 C_F (-4 +   ~d)^2   ~d (-479744 + 1418752   ~d 
- 1664968   ~d^2 + 989740   ~d^3 
\nonumber\\[2ex]
&&
- 297578   ~d^4 
+ 28179   ~d^5 + 7786   ~d^6 
- 2351   ~d^7 + 184   ~d^8) n_f 
+ C_A (-3440640 
\nonumber\\[2ex]
&& + 18589696   ~d - 42184960   ~d^2 
+ 55760640   ~d^3 - 
         47369168   ~d^4 + 26855488   ~d^5 
\nonumber\\[2ex]
&&
- 10323440   ~d^6 
+ 2698144   ~d^7 - 
         474715   ~d^8 + 54662   ~d^9 - 3837   ~d^{10} 
\nonumber\\[2ex]
&&+ 130   ~d^{11}) n_f\Bigg\}\Bigg)\Bigg]\Bigg/
 \Bigg[8 (-4 +   ~d)^3 (-3 +   ~d) (-2 +   ~d)^2 (-1 +   ~d)   ~d (-7 + 2   ~d) 
\nonumber\\[2ex]
&&\times (-5 + 2   ~d) (-8 + 3   ~d) (2 d^2 - 3 d - 8) \Bigg] \,,
\end{eqnarray}
where the color factor $C_F=(N^2-1)/2 N$.
For the quark form factor, we obtain $\hat F^{T,(0)}_q=1$  and 
\begin{eqnarray}
\hat F^{T,(1)}_q&=&
2 \Bigg[-{i\over 2}~ A_{2,LO} ~ C_F ~ \Big(64 - 34 ~ d + 5 ~ d^2\Big)\Bigg]\Bigg/
\Big[(d -4) (d-2) \Big] \, ,
\\[2ex]
\hat F^{T,(2)}_q&=&
-\Bigg[C_F ~ \Bigg(16 ~ A_{2,LO}^2 ~ \Big(-3360 + 5524 ~ d - 3607 ~ d^2 + 1169 ~ d^3 - 188 ~ d^4 
+ 12 ~ d^5\Big) ~ 
\nonumber\\[2ex]
&& \times     \Bigg\{C_F ~ d ~ \Big(2048 - 5312 ~ d + 5156 ~ d^2 - 2432 ~ d^3 + 619 ~ d^4 
- 84 ~ d^5 + 5 ~ d^6\Big) 
\nonumber\\[2ex]
&&
+ 16 ~ C_A ~ \Big(192 - 288 ~ d - 28 ~ d^2 + 322 ~ d^3 - 255 ~ d^4 
+ 89 ~ d^5 - 15 ~ d^6 + d^7\Big) 
\nonumber\\[2ex]
&&
+ 4 ~ \Big(-4 + d\Big)^2 ~ \Big(48 - 56 ~ d + 35 ~ d^2 - 13 ~ d^3 
+ 2 ~ d^4\Big) ~ n_f\Bigg\} + 
    A_3 ~ \Big(-8 + 3 ~ d\Big) 
\nonumber\\[2ex]
&&
~\times \Bigg\{C_A ~ \Big(6881280 
- 11370496 ~ d 
- 5231104 ~ d^2 + 24600896 ~ d^3 - 18172384 ~ d^4 
\nonumber\\[2ex]
&&
- 2105928 ~ d^5 
+ 11581460 ~ d^6 - 8688682 ~ d^7 + 
        3558513 ~ d^8 - 910210 ~ d^9 
\nonumber\\[2ex]
&&+ 146166 ~ d^{10} - 13592 ~ d^{11} + 561 ~ d^{12}\Big) - 
      2 ~ \Big(C_F ~ d ~ \Big(-13110272 
+ 59524736 ~ d 
\nonumber\\[2ex]
&&
- 119256896 ~ d^2 + 139910176 ~ d^3 - 
          107260248 ~ d^4 
+ 56580636 ~ d^5 - 20992430 ~ d^6 
\nonumber\\[2ex]
&&
+ 5484477 ~ d^7 - 
          989746 ~ d^8 
+ 117582 ~ d^9 - 8276 ~ d^{10} + 261 ~ d^{11}\Big) - 
        16 ~ \Big(-4 + d\Big)^2 ~
\nonumber\\[2ex]
&&
\times \Big(13440 
- 1936 ~ d - 63236 ~ d^2 + 124494 ~ d^3 - 119835 ~ d^4 + 
          68959 ~ d^5 
\nonumber\\[2ex]
&&- 24789 ~ d^6 + 5463 ~ d^7 - 676 ~ d^8 + 36 ~ d^9\Big) ~ n_f\Big)\Bigg\} + 
    2 ~ \Big(-12 + 19 ~ d 
- 8 ~ d^2 + d^3\Big) ~ 
\nonumber\\[2ex]
&&
\times \Bigg(2 ~ A_6 ~ \Big(-4 + d\Big)^2 ~ d ~ \Big(-16 + 30 ~ d - 17 ~ d^2 + 
        3 ~ d^3\Big) ~
\Bigg\{2 ~ C_F ~ \Big(-288 + 192 ~ d 
\nonumber\\[2ex]
&& 
- 16 ~ d^2 - 6 ~ d^3 + d^4\Big) + 
        C_A ~ \Big(416 - 192 ~ d 
+ 24 ~ d^2 - 14 ~ d^3 + 3 ~ d^4\Big)\Bigg\} 
\nonumber\\[2ex]
&&
+ A_4 ~ \Bigg\{C_A ~ \Big(6881280 - 23730176 ~ d 
+ 37309440 ~ d^2 - 32039680 ~ d^3 
\nonumber\\[2ex]
&&
+ 15239072 ~ d^4 
- 3272584 ~ d^5 
- 345324 ~ d^6 + 415114 ~ d^7 - 116931 ~ d^8 + 
          17142 ~ d^9 
\nonumber\\[2ex]
&&
- 1361 ~ d^{10} + 48 ~ d^{11}\Big) 
- 2 ~ \Big(C_F ~ d ~ \Big(677888 - 2026112 ~ d + 2909696 ~ d^2 
\nonumber\\[2ex]
&&
- 2895040 ~ d^3 
+ 2126552 ~ d^4 - 1101532 ~ d^5 + 384546 ~ d^6 - 87351 ~ d^7 + 12286 ~ d^8 
\nonumber\\[2ex]
&&- 965 ~ d^9 + 32 ~ d^{10}\Big) - 4 ~ \Big(860160 - 1270784 ~ d + 218048 ~ d^2 + 
            766736 ~ d^3 
\nonumber\\[2ex]
&&- 743952 ~ d^4 + 330352 ~ d^5 - 81952 ~ d^6 + 10967 ~ d^7 - 
            533 ~ d^8 - 36 ~ d^9 
\nonumber\\[2ex]
&&+ 4 ~ d^{10}\Big) ~ n_f\Big)\Bigg\}\Bigg)\Bigg)\Bigg]\Bigg/
 \Big[16 ~ \Big(-4 + d\Big)^3 ~ \Big(-3 + d\Big) ~ \Big(-2 + d\Big)^2 ~ 
 \Big(-1 + d\Big)^2 ~ d ~ 
\nonumber\\[2ex]
&&\times \Big(-7 + 2 ~ d\Big) ~ \Big(-5 + 2 ~ d\Big) ~ 
  \Big(-8 + 3 ~ d\Big)\Big]
\, .
\end{eqnarray}
The exact results for the master integrals $A_i$ ($i=\{2,LO\},3,4,6$) can
be expressed in terms of Euler Gamma functions and are available in the
works on two-loop electroweak form factors \cite{Gehrmann:2005pd}.  The
most difficult crossed two-loop master
integral $A_6$ was solved exactly in \cite{Gehrmann:2005pd}.  These results
are used to present the form factors to order ${\cal O}(\varepsilon^4)$ and
are given in the appendix.  We use them to study the infrared pole structure
of these farm factors in the next section.  The higher order terms ${\cal O}
(\varepsilon^i), i>0$ are also useful to perform ultraviolet renormalisation
of the form factors beyond two-loop level.  

\section{Infrared divergence structure}
Having obtained these form factors at two-loop level, the next step is to
study the infrared pole structure of these factors in order to establish the
universal behaviour of these QCD amplitudes.  In the past, there have been 
detailed studies of quark and gluon form factors
through Sudakov integro-differential equation \cite{Sudakov:1954sw,Mueller:1979ih, 
Collins:1980ih,Sen:1981sd,col}, see also 
\cite{Ravindran:2004mb,Moch:2005ky,Laenen:2005uz,Idilbi:2005ni,Ravindran:2005vv}.    
Since the KK modes are colour
singlet fields, the unrenormalised form factors $\hat F^T_I(\hat a_s,Q^2,\mu^2,\varepsilon)$ 
are expected to satisfy
similar integro-differential equation that follows from the gauge 
as well as renormalisation group (RG) invariances.
In dimensional regularisation,
\begin{eqnarray}
Q^2{d \over dQ^2} \ln \hat F_I^T\left(\hat a_s,Q^2,\mu^2,\varepsilon\right)&=&
{1 \over 2 }
\Bigg[K^{T,I}\left(\hat a_s,{\mu_R^2 \over \mu^2},\varepsilon\right)  
+ G^{T,I}\left(\hat a_s,{Q^2 \over \mu_R^2},{\mu_R^2 \over \mu^2},\varepsilon\right)
\Bigg] \,,
\label{sud1}
\end{eqnarray}
where the constants $K^{T,I}$ contain all the poles in $\varepsilon$,  
and $G^{T,I}$ are finite as $\varepsilon$ becomes zero.  
The RG invariance of $\hat F_I^T$ gives
\begin{eqnarray}
\mu_R^2 {d \over d\mu_R^2} 
K^{T,I}\Bigg(\hat a_s,{\mu_R^2 \over \mu^2},\varepsilon\Bigg)=-A^{T,I}(a_s(\mu_R^2))\,,
\nonumber\\[2ex]
\mu_R^2 {d \over d\mu_R^2} 
G^{T,I}\Bigg(\hat a_s,{Q^2\over \mu_R^2},
{\mu_R^2 \over \mu^2},\varepsilon\Bigg)=A^{T,I}(a_s(\mu_R^2)) \, .
\end{eqnarray}
The quantities $A^{T,I}$ are the cusp anomalous dimensions which  
are expanded as
\begin{eqnarray}
A^{T,I}(a_s(\mu_R^2))=\sum_{i=1}^\infty a_s^{i}(\mu_R^2)~ A_i^{T,I}.\,
\end{eqnarray} 
Solving these RG equations, the constants $K^{T,I}$ and $G^{T,I}$ can be obtained in powers of bare coupling constant
$\hat a_s$.  Using these solutions, we obtain, 
\begin{eqnarray}
\ln \hat F_I^T(\hat a_s,Q^2,\mu^2,\varepsilon)
&=&\sum_{i=1}^\infty \hat a_s^i 
\left({Q^2 \over \mu^2}\right)^{i {\varepsilon \over 2}}S^i_{\varepsilon}~ 
\hat {\cal L}_{F^T}^{I(i)}(\varepsilon) \,,
\label{logF}
\end{eqnarray}
where
\begin{eqnarray}
\hat {\cal L}_{F^T}^{I(1)}&=&{1\over \varepsilon^2} \Bigg(-2 A_1^{T,I}\Bigg) 
              +{1 \over \varepsilon} \Bigg(G_1^{T,I}(\varepsilon)\Bigg) \,,
\nonumber\\[2ex]
\hat {\cal L}_{F^T}^{I(2)}&=&{1\over \varepsilon^3} \Bigg(\beta_0 A_1^{T,I}\Bigg) 
                  +{1\over \varepsilon^2} \Bigg(-{1 \over 2} A_2^{T,I} 
                  - \beta_0  G_1^{T,I}(\varepsilon)\Bigg)
                  +{1 \over 2 \varepsilon} G_2^{T,I}(\varepsilon) \,.
\label{curlL}
\end{eqnarray}
The cusp anomalous dimensions $A^{T,I}_i$
can be obtained by comparing eqns.(\ref{logF},\ref{curlL}) and the results of the
form factors, eqns(\ref{fnlog},\ref{fnnlog},\ref{fnloq},\ref{fnnloq}).  We find that
they are identical to those obtained in \cite{Kodaira:1981nh}, that
is, those appearing in gluon and quark form factors, confirming the 
universality of IR structure of these form factors. 
The coefficients $G^{T, I}_i(\varepsilon)$ take the following form 
\begin{eqnarray}
G^{T,I}_1(\varepsilon)&=& 
2~B^{T,I}_1  + f_1^{T,I}
        +\sum_{k=1}^\infty \varepsilon^k  g^{~{T,I},k}_1\,,
\nonumber \\[2ex]
G^{{T,I}}_2(\varepsilon)&=& 
2~B_2^{T,I} + f_2^{T,I} 
        -2 \beta_0  g^{~{T,I},1}_1
        +\sum_{k=1}^\infty \varepsilon^k  g^{~{T,I},k}_2\,,
\end{eqnarray}
where again the collinear anomalous dimension $B_i^{T,I}$ and soft anomalous dimension 
$f_i^{T,I}$ are found to be identical to 
$B_i^I$ and $f_i^I$ obtained in \cite{moch:2004mvv,Ravindran:2004mb} for quark and 
gluon form factors.  We find that only $g^{~{T,I},k}_i$ are operator dependent.

Another independent check on our computation is done by establishing the connection 
between these form factors and the very successful proposal by 
Catani \cite{catani} (also see \cite{sterman}) 
on one and two-loop QCD amplitudes using the universal 
factors $\mathbf{I}^{(i)}_I(\varepsilon)$ and $\mathbf{H}^{(i)}_I$, $i=1,2$. 
The all order generalisation of Catani's proposal was obtained by Becher and Neubert \cite{Becher:2009cu}
and also by Gardi and Magnea \cite{Gardi:2009qi}.
  These
universal factors capture all the IR poles
of n-parton QCD amplitudes up to two-loop level in QCD. 
Following \cite{catani}, we proceed by expressing 
the matrix elements in terms of  UV renormalised ones as 
\begin{eqnarray}
\mathbf{\hat {\cal M}}_I =\mathbf{{M}}_I^{(0)} + a_s(\mu_R^2)
\mathbf{ {M}}_I^{(1)}
+ a_s^2(\mu_R^2)
\mathbf{ {M}}_I^{(2)} + {\cal O}\left(a_s^3(\mu_R^2)\right)
\,,\, \quad \quad I=g,q,\overline q \,.
\end{eqnarray}
Using the universal $\mathbf{I}_I(\varepsilon)$ obtained by Catani, we can write down
\begin{eqnarray}
\mathbf{M}_I^{(1)}&=&2 \mathbf{I}^{(1)}_I(\varepsilon) \mathbf{M}^{(0)}_I(\varepsilon)
+\mathbf{M}^{(1)}_{I,fin}(\varepsilon) \, ,
\nonumber\\[2ex]
\mathbf{M}_I^{(2)}&=&2 \mathbf{I}^{(1)}_I(\varepsilon) \mathbf{M}^{(1)}_I(\varepsilon)
+4 \mathbf{I}^{(2)}_I(\varepsilon) \mathbf{M}^{(0)}_I(\varepsilon) 
+\mathbf{M}^{(2)}_{I,fin}(\varepsilon) \, .
\end{eqnarray}
In terms of these $\mathbf{M}_I^{(i)}$, we find
\begin{eqnarray}
\hat F_I^{T,(1)}&=&2 \mu_R^\varepsilon 
{{\mathbf{I}}}_I^{(1)}(\varepsilon)
+\hat F_{I,fin}^{T,(1)}(\varepsilon) \, ,
\nonumber\\[2ex]
\hat F_I^{T,(2)}&=& 4 \mu_R^{2 \varepsilon}
\Bigg[ \left(\mathbf{I}^{(1)}_I(\varepsilon)\right)^2
      +\mathbf{I}^{(2)}_I(\varepsilon)
      -{\beta_0 \over \varepsilon} \Big(\mathbf{I}^{(1)}_I(\varepsilon)
        +{\mu_R^{-\varepsilon} \over 2} \hat F^{T,(1)}_{I,fin}(\varepsilon)\Big)
\nonumber\\[2ex]
&&        +{1 \over 2} \mu_R^{-\varepsilon} \mathbf{I}^{(1)}_I(\varepsilon)
         \hat F^{T,(1)}_{I,fin}(\varepsilon)\Bigg] 
        +\hat F^{T,(2)}_{I,fin}(\varepsilon) \, ,
\end{eqnarray}
where 
\begin{eqnarray}
\hat F^{T,(i)}_{I,fin}(\varepsilon) &=& \mu_R^{i \varepsilon}
{\mathbf{M}^{(0)*}_I\cdot \mathbf{M}^{(i)}_{I,fin} \over {\mathbf{M}^{(0)*}_I}
\cdot \mathbf{M}^{(0)}_I} \, ,
\quad \quad \quad i=1,2 \, .
\end{eqnarray}
The singular universal functions $\mathbf{I}_I^{(i)}$ are given by
\begin{eqnarray}
\mathbf{I}^{(1)}_q(\varepsilon)&=&-{e^{-\varepsilon \gamma_E/2} \over
                           \Gamma\left(1+{\varepsilon\over 2}\right)}
\left( {Q^2 \over \mu_R^2}\right)^{{\varepsilon \over 2}}
\left(4{ C_F \over \varepsilon^2} - 3 { C_F \over \varepsilon}\right) \,,
\nonumber\\[2ex]
\mathbf{I}^{(1)}_g(\varepsilon)&=&-{e^{-\varepsilon \gamma_E/2} \over
                           \Gamma\left(1+{\varepsilon\over 2}\right)}
\left( {Q^2 \over \mu_R^2}\right)^{{\varepsilon \over 2}}
\left(4{ C_A \over \varepsilon^2} - { \beta_0 \over \varepsilon}\right) \,,
\\[2ex]
\mathbf{I}^{(2)}_I(\varepsilon)&=&-{1 \over 2} \left(\mathbf{I}^{(1)}_I(\varepsilon)\right)^2
+ {\beta_0 \over \varepsilon}\mathbf{I}^{(1)}_I(\varepsilon) 
\nonumber\\[2ex]
&&+ {e^{{\varepsilon \gamma_E \over 2}} \Gamma\left(1+\varepsilon\right)
\over \Gamma\left(1+{\varepsilon \over 2}\right)}
\left(-{\beta_0\over \varepsilon} + K \right) \mathbf{I}^{(1)}_I(2 \varepsilon)
+\mathbf{H}^{(2)}_I {1 \over \varepsilon}  \, ,
\end{eqnarray}
and
\begin{eqnarray}
K=\left({67 \over 18} - \zeta_2\right) C_A -{10 \over 9} T_F n_f \,.
\end{eqnarray}
Using our results for $\hat F^{T,(i)}$ given in eqns.(\ref{fnlog},\ref{fnnlog},\ref{fnloq},\ref{fnnloq}) and the results for $\mathbf{I}_I^{(i)}$ given in
\cite{catani}, we obtain $\mathbf{H}_I^{(2)}$: 
\begin{eqnarray}
\mathbf{H}_g^{(2)}&=&C_A^2 \Bigg(-{5 \over 12} -{11 \over 24} \zeta_2 -{1 \over 2} \zeta_3\Bigg)
         +C_A n_f \Bigg({29 \over 27} +{1 \over 12} \zeta_2\Bigg)
+C_F n_f \Bigg(-{1 \over 2}\Bigg)
         +n_f^2 \Bigg(-{5 \over 27}\Bigg) \, ,
\nonumber\\[2ex]
\mathbf{H}_q^{(2)}&=&C_F^2 \Bigg({3 \over 8} -3 \zeta_2 + 6 \zeta_3\Bigg)
          +C_A C_F \Bigg(-{245 \over 216} +{23 \over 8} \zeta_2-{13 \over 2} \zeta_3\Bigg)
+C_F n_f \Bigg({25 \over 108} -{1 \over 4} \zeta_2\Bigg) \, .
\end{eqnarray}
The single pole coefficients thus obtained agree with the color diagonal part of eqn.(12) of
\cite{Becher:2009cu} (see also eqn.(4.21) of \cite{Ravindran:2004mb} for quark and gluon form
factors and \cite{Anastasiou:2001sv,Anastasiou:2000ue,Anastasiou:2000kg,Glover:2001af} for
four parton amplitudes).  This serves as a check on our computation and also establishes the
proposal by Catani on IR universality of QCD amplitudes with $T_{\mu\nu}$ insertion.
%
%
%

\section{Conclusions}

We present an important ingredient to the full NNLO QCD correction to graviton
mediated hadronic scattering processes namely the gluon and quark form factors
of energy momentum tensor of the QCD part of the SM up to two-loop level in QCD.
We have used dimensional regularisation to obtain these form factors in $SU(N)$
gauge theory with $n_f$ light flavours.  Both exact as well as expanded results
in $\varepsilon$ are presented.  The higher order terms in $\varepsilon$ of these
form factors are important for the ultraviolet renormalisation of these amplitudes
at three-loop level.  We have shown that these form factors satisfy Sudakov
integro-differential equation with same cusp $A_I$, collinear $B^I$ and soft
$f^I$ anomalous dimensions that contribute to electroweak vector boson and gluon
form factors.  In addition, they also show the universal behaviour of the infrared
poles in $\varepsilon$ in accordance with the proposal by Catani.     

Spin-2 resonance production has been widely studied in the context of the 
Higgs \cite{HC} and BSM models \cite{A_CE}.  The two-loop results presented in
this paper would further reduce the theoretical uncertainties and hence
improve the predictions in disentangling the various postulates.  We further
plan to apply these two-loop results to the TeV scale gravity models \cite{future}.

\section*{Acknowledgements}

The work of DdeF and JM was supported in part by UBACYT, CONICET, ANPCyT and the
Research Executive Agency (REA) of the European Union under the Grant Agreement
number PITN-GA-2010-264564 (LHCPhenoNet).  MM would like to thank IMSc for hospitality.

\section*{Appendix}

We present here the form factors as a series expansion in $\varepsilon$ up to
${\cal O}(\varepsilon^4)$ for $F_I^{(1)}$ and up to ${\cal O}(\varepsilon^2))$ 
for $F_I^{(2)}$:

\begin{eqnarray}
   \hat F^{T,(1)}_g &=&
        n_f \Bigg[
        {1 \over  \varepsilon }     \Bigg(
          - {4  \over  3}
          \Bigg)
       +     \Bigg(
           {35  \over  18}
          \Bigg)
       +  \varepsilon       \Bigg(
          - {497  \over  216}
          + {1  \over  6}    \zeta_2
          \Bigg)
       +  \varepsilon ^2      \Bigg(
           {6593  \over  2592}
          - {7  \over  18}    \zeta_3
          - {35  \over  144}    \zeta_2
          \Bigg)
\nonumber\\[2ex]
&&       +  \varepsilon ^3      \Bigg(
          - {84797  \over  31104}
          + {245  \over  432}    \zeta_3
          + {497  \over  1728}    \zeta_2
          + {47  \over  480}    \zeta_2^2
          \Bigg)
       +  \varepsilon ^4      \Bigg(
           {1072433  \over  373248}
          - {31  \over  120}    \zeta_5
\nonumber\\[2ex]
&&          - {3479  \over  5184}    \zeta_3
          - {6593  \over  20736}    \zeta_2
          + {7  \over  144}    \zeta_2    \zeta_3
          - {329  \over  2304}    \zeta_2^2
          \Bigg)
         \Bigg]
       +C_A \Bigg[
         {1 \over  \varepsilon ^2 }   \Bigg(
          - 8
          \Bigg)
       +   {1 \over  \varepsilon}    \Bigg(
           {22  \over  3}
          \Bigg)
\nonumber\\[2ex]
&&       +     \Bigg(
          - {203  \over  18}
          +   \zeta_2
          \Bigg)
       +    \varepsilon     \Bigg(
           {2879  \over  216}
          - {7  \over  3 }   \zeta_3
          - {11  \over  12 }   \zeta_2
          \Bigg)
       +    \varepsilon ^2    \Bigg(
          - {37307  \over  2592}
          + {77  \over  36}    \zeta_3
          + {203  \over  144}    \zeta_2
\nonumber\\[2ex]
&&          + {47  \over  80}    \zeta_2^2
          \Bigg)
       +    \varepsilon ^3    \Bigg(
           {465143  \over  31104}
          - {31  \over  20}    \zeta_5
          - {1421  \over  432}    \zeta_3
          - {2879  \over  1728}    \zeta_2
          + {7  \over  24}    \zeta_2    \zeta_3
          - {517  \over  960}    \zeta_2^2
          \Bigg)
\nonumber\\[2ex]
&&       +    \varepsilon ^4    \Bigg(
          - {5695811  \over  373248}
          + {341  \over  240}    \zeta_5
          + {20153  \over  5184}    \zeta_3
          - {49  \over  144}    \zeta_3^2
          + {37307  \over  20736}    \zeta_2
          - {77  \over  288}    \zeta_2    \zeta_3
\nonumber\\[2ex]
&&          + {9541  \over  11520}    \zeta_2^2
          + {949  \over  4480}    \zeta_2^3
          \Bigg)
          \Bigg]
\label{fnlog}
\\[2ex]
   \hat F^{T,(2)}_g &=&
        C_F n_f\Bigg[
          {1 \over  \varepsilon  }     \Bigg(
          - 2
          \Bigg)
       +     \Bigg(
           {61  \over  6}
          - 8    \zeta_3
          \Bigg)
       +    \varepsilon       \Bigg(
          - {2245  \over  72}
          + {59  \over  3}    \zeta_3
          + {1  \over  2}    \zeta_2
          + {12  \over  5}    \zeta_2^2
          \Bigg)
\nonumber\\[2ex]
&&       +   \varepsilon ^2     \Bigg(
           {64177  \over  864}
          - {14    \zeta_5}
          - {335  \over  9}    \zeta_3
          - {83  \over  24}    \zeta_2
          + 2    \zeta_2    \zeta_3
          - {179  \over  30}    \zeta_2^2
          \Bigg)
       +C_A n_f \Bigg[
         {1 \over  \varepsilon ^3}      \Bigg(
           8
          \Bigg)
\nonumber\\[2ex]
&&       +   {1 \over  \varepsilon ^2}      \Bigg(
          - {40  \over  3}
          \Bigg)
       +   {1 \over  \varepsilon }      \Bigg(
           {41  \over  3}
          - {2  \over  3}    \zeta_2
          \Bigg)
       +     \Bigg(
          - {605  \over  108}
          + {10    \zeta_3}
          + {5  \over  9}    \zeta_2
          \Bigg)
       +   \varepsilon      \Bigg(
          - {21557  \over  1296}
\nonumber\\[2ex]
&&          - {182  \over  9}    \zeta_3
          + {145  \over  108}    \zeta_2
          - {57  \over  20}    \zeta_2^2
          \Bigg)
       +    \varepsilon ^2      \Bigg(
           {320813  \over  5184}
          + {71  \over  10}    \zeta_5
          + {6407  \over  216}    \zeta_3
          - {3617  \over  648}    \zeta_2
\nonumber\\[2ex]
&&          - {43  \over  18}    \zeta_2    \zeta_3
          + {1099  \over  180}    \zeta_2^2
          \Bigg)
  +C_A^2  \Bigg[
         {1 \over  \varepsilon ^4}    \Bigg(
           32
          \Bigg)
       +   {1 \over  \varepsilon ^3}    \Bigg(
          - 44
          \Bigg)
       +   {1 \over  \varepsilon ^2}    \Bigg(
           {226  \over  3}
          - 4    \zeta_2
          \Bigg)
\nonumber\\[2ex]
&&       +   {1 \over  \varepsilon  }   \Bigg(
          - 81
          + {50  \over  3}    \zeta_3
          + {11  \over  3}    \zeta_2
          \Bigg)
       +     \Bigg(
           {5249  \over  108}
          - {11    \zeta_3}
          - {67  \over  18}    \zeta_2
          - {21  \over  5}    \zeta_2^2
          \Bigg)
       +    \varepsilon     \Bigg(
           {59009  \over  1296}
\nonumber\\[2ex]
&&          - {71  \over  10}    \zeta_5
          + {433  \over  18}    \zeta_3
          - {337  \over  108}    \zeta_2
          - {23  \over  6}    \zeta_2    \zeta_3
          + {99  \over  40}    \zeta_2^2
          \Bigg)
       +    \varepsilon ^2    \Bigg(
          - {1233397  \over  5184}
          + {759  \over  20}    \zeta_5
\nonumber\\[2ex]
&&          - {8855  \over  216}    \zeta_3
          + {901  \over  36}    \zeta_3^2
          + {12551  \over  648}    \zeta_2
          + {77  \over  36}    \zeta_2    \zeta_3
          - {4843  \over  720}    \zeta_2^2
          + {2313  \over  280}    \zeta_2^3
          \Bigg)
        \Bigg]
\label{fnnlog}
\\[2ex]
   \hat F^{T,(1)}_q &=&
    C_F \Bigg[
         {1 \over \varepsilon^2}   \Bigg(
          - 8
          \Bigg)
       +  {1 \over \varepsilon}   \Bigg(
           6
          \Bigg)
       +    \Bigg(
          - 10
          + \zeta_2
          \Bigg)
       +  \varepsilon   \Bigg(
           12
          - {7  \over  3} \zeta_3
          - {3  \over  4} \zeta_2
          \Bigg)
\nonumber\\[2ex]
&&       +  \varepsilon^2   \Bigg(
          - 13
          + {7  \over  4} \zeta_3
          + {5  \over  4} \zeta_2
          + {47  \over  80} \zeta_2^2
          \Bigg)
       +  \varepsilon^3   \Bigg(
           {27  \over  2}
          - {31  \over  20} \zeta_5
          - {35  \over  12} \zeta_3
          - {3  \over  2} \zeta_2
\nonumber\\[2ex]
&&          + {7  \over  24} \zeta_2 \zeta_3
          - {141  \over  320} \zeta_2^2
          \Bigg)
       +  \varepsilon^4   \Bigg(
          - {55  \over  4}
          + {93  \over  80} \zeta_5
          + {7  \over  2} \zeta_3
          - {49  \over  144} \zeta_3^2
          + {13  \over  8} \zeta_2
\nonumber\\[2ex]
&&          - {7  \over  32} \zeta_2 \zeta_3
          + {47  \over  64} \zeta_2^2
          + {949  \over  4480} \zeta_2^3
          \Bigg)
      \Bigg]
\label{fnloq}
\end{eqnarray}
\begin{eqnarray}
   \hat F^{T,(2)}_q &=&
      C_F n_f \Bigg[
         {1 \over \varepsilon^3}    \Bigg(
          - {8  \over  3}
          \Bigg)
       + {1 \over \varepsilon^2}    \Bigg(
           {56  \over  9}
          \Bigg)
       +  {1 \over \varepsilon}    \Bigg(
          - {425  \over  27}
          - {2  \over  3} \zeta_2
          \Bigg)
       +    \Bigg(
           {9989  \over  324}
          - {26  \over  9} \zeta_3
          + {38  \over  9} \zeta_2
          \Bigg)
\nonumber\\[2ex]
&&       +  \varepsilon    \Bigg(
          - {202253  \over  3888}
          + {2  \over  27} \zeta_3
          - {989  \over  108} \zeta_2
          + {41  \over  60} \zeta_2^2
          \Bigg)
       +  \varepsilon^2    \Bigg(
           {3788165  \over  46656}
          - {121  \over  30} \zeta_5
          - {935  \over  324} \zeta_3
\nonumber\\[2ex]
&&          + {22937  \over  1296} \zeta_2
          - {13  \over  18} \zeta_2 \zeta_3
          + {97  \over  180} \zeta_2^2
          \Bigg)
     \Bigg]
     +C_F^2 \Bigg[
         {1 \over \varepsilon^4}   \Bigg(
           32
          \Bigg)
       +  {1 \over \varepsilon^3}   \Bigg(
          - 48
          \Bigg)
\nonumber\\[2ex]
&&      +  {1 \over \varepsilon^2}   \Bigg(
           98
          - 8 \zeta_2
          \Bigg)
       +  {1 \over \varepsilon}  \Bigg(
          - {309  \over  2}
          + {128  \over  3} \zeta_3
          \Bigg)
       +   \Bigg(
           {5317  \over  24}
          - 90 \zeta_3
          + {41  \over  2} \zeta_2
          - 13 \zeta_2^2
          \Bigg)
\nonumber\\[2ex]
&&       +  \varepsilon   \Bigg(
          - {28127  \over  96}
          + {92  \over  5} \zeta_5
          + {1327  \over  6} \zeta_3
          - {1495  \over  24} \zeta_2
          - {56  \over  3} \zeta_2 \zeta_3
          + {173  \over  6} \zeta_2^2
          \Bigg)
\nonumber\\[2ex]
&&       +  \varepsilon^2   \Bigg(
           {1244293  \over  3456}
          - {311  \over  10} \zeta_5
          - {34735  \over  72} \zeta_3
          + {652  \over  9} \zeta_3^2
          + {38543  \over  288} \zeta_2
          + {193  \over  6} \zeta_2 \zeta_3
          - {10085  \over  144} \zeta_2^2
\nonumber\\[2ex]
&&          + {223  \over  20} \zeta_2^3
          \Bigg)
         \Bigg]
+    C_A C_F \Bigg[
         {1 \over \varepsilon^3}   \Bigg(
           {44  \over  3}
          \Bigg)
       +  {1 \over \varepsilon^2}   \Bigg(
          - {332  \over  9}
          + 4 \zeta_2
          \Bigg)
       +  {1 \over \varepsilon}   \Bigg(
           {4921  \over  54}
          - 26 \zeta_3
\nonumber\\[2ex]
&&          + {11  \over  3} \zeta_2
          \Bigg)
       +    \Bigg(
          - {120205  \over  648}
          + {755  \over  9} \zeta_3
          - {251  \over  9} \zeta_2
          + {44  \over  5} \zeta_2^2
          \Bigg)
       +  \varepsilon   \Bigg(
           {2562925  \over  7776}
          - {51  \over  2} \zeta_5
\nonumber\\[2ex]
&&          - {5273  \over  27} \zeta_3
          + {14761  \over  216} \zeta_2
          + {89  \over  6} \zeta_2 \zeta_3
          - {3299  \over  120} \zeta_2^2
          \Bigg)
       +  \varepsilon^2   \Bigg(
          - {50471413  \over  93312}
          + {3971  \over  60} \zeta_5
\nonumber\\[2ex]
&&          + {282817  \over  648} \zeta_3
          - {569  \over  12} \zeta_3^2
          - {351733  \over  2592} \zeta_2
          - {1069  \over  36} \zeta_2 \zeta_3
          + {7481  \over  120} \zeta_2^2
          - {809  \over  280} \zeta_2^3
          \Bigg)
        \Bigg]
\label{fnnloq}
\end{eqnarray}

\end{document}